\documentclass[twocolumn,superscriptaddress,amsmath,amssymb,prx,floatfix]{revtex4-2}

\usepackage{xr}
\usepackage{hyperref}
\hypersetup{
    colorlinks=true,
    linkcolor=blue,
    filecolor=magenta,      
    urlcolor=cyan,
    pdfpagemode=FullScreen,
    }

\usepackage[belowskip=-10pt,aboveskip=10pt,font=small,justification=Justified,format=plain]{caption}

\usepackage{threeparttable}

\usepackage{graphicx}
\usepackage{dcolumn}
\usepackage{bm}

\usepackage[utf8]{inputenc}
\usepackage[T1]{fontenc}
\usepackage{amsmath}
\usepackage{mathptmx}
\usepackage[dvipsnames]{xcolor}
\usepackage{geometry}
\usepackage[explicit]{titlesec}

\usepackage{xcolor}
\usepackage[draft]{changes}
\definechangesauthor[color=red, name={Paulo Santos}]{PVS}
\definechangesauthor[color=cyan, name={Alexander Kuznetsov}]{AK}

\usepackage{verbatim}
\setauthormarkup{}

\RequirePackage[normalem]{ulem} 
\RequirePackage{color}\definecolor{RED}{rgb}{1,0,0}\definecolor{BLUE}{rgb}{0,0,1} 
\providecommand{\DIFaddbegin}{} 
\providecommand{\DIFaddend}{} 
\providecommand{\DIFdelbegin}{} 
\providecommand{\DIFdelend}{} 
\providecommand{\DIFaddbeginFL}{} 
\providecommand{\DIFaddendFL}{} 
\providecommand{\DIFdelbeginFL}{} 
\providecommand{\DIFdelendFL}{} 
\newcommand{\DIFscaledelfig}{0.5}
\RequirePackage{settobox} 
\RequirePackage{letltxmacro} 
\newsavebox{\DIFdelgraphicsbox} 
\newlength{\DIFdelgraphicswidth} 
\newlength{\DIFdelgraphicsheight} 
\LetLtxMacro{\DIFOincludegraphics}{\includegraphics} 
\newcommand{\DIFaddincludegraphics}[2][]{{\color{blue}\fbox{\DIFOincludegraphics[#1]{#2}}}} 
\newcommand{\DIFdelincludegraphics}[2][]{
\sbox{\DIFdelgraphicsbox}{\DIFOincludegraphics[#1]{#2}}
\settoboxwidth{\DIFdelgraphicswidth}{\DIFdelgraphicsbox} 
\settoboxtotalheight{\DIFdelgraphicsheight}{\DIFdelgraphicsbox} 
\scalebox{\DIFscaledelfig}{
\parbox[b]{\DIFdelgraphicswidth}{\usebox{\DIFdelgraphicsbox}\\[-\baselineskip] \rule{\DIFdelgraphicswidth}{0em}}\llap{\resizebox{\DIFdelgraphicswidth}{\DIFdelgraphicsheight}{
\setlength{\unitlength}{\DIFdelgraphicswidth}
\begin{picture}(1,1)
\thicklines\linethickness{2pt} 
{\color[rgb]{1,0,0}\put(0,0){\framebox(1,1){}}}
{\color[rgb]{1,0,0}\put(0,0){\line( 1,1){1}}}
{\color[rgb]{1,0,0}\put(0,1){\line(1,-1){1}}}
\end{picture}
}\hspace*{3pt}}} 
} 
\LetLtxMacro{\DIFOaddbegin}{\DIFaddbegin} 
\LetLtxMacro{\DIFOaddend}{\DIFaddend} 
\LetLtxMacro{\DIFOdelbegin}{\DIFdelbegin} 
\LetLtxMacro{\DIFOdelend}{\DIFdelend} 
\DeclareRobustCommand{\DIFaddbegin}{\DIFOaddbegin \let\includegraphics\DIFaddincludegraphics} 
\DeclareRobustCommand{\DIFaddend}{\DIFOaddend \let\includegraphics\DIFOincludegraphics} 
\DeclareRobustCommand{\DIFdelbegin}{\DIFOdelbegin \let\includegraphics\DIFdelincludegraphics} 
\DeclareRobustCommand{\DIFdelend}{\DIFOaddend \let\includegraphics\DIFOincludegraphics} 
\LetLtxMacro{\DIFOaddbeginFL}{\DIFaddbeginFL} 
\LetLtxMacro{\DIFOaddendFL}{\DIFaddendFL} 
\LetLtxMacro{\DIFOdelbeginFL}{\DIFdelbeginFL} 
\LetLtxMacro{\DIFOdelendFL}{\DIFdelendFL} 
\DeclareRobustCommand{\DIFaddbeginFL}{\DIFOaddbeginFL \let\includegraphics\DIFaddincludegraphics} 
\DeclareRobustCommand{\DIFaddendFL}{\DIFOaddendFL \let\includegraphics\DIFOincludegraphics} 
\DeclareRobustCommand{\DIFdelbeginFL}{\DIFOdelbeginFL \let\includegraphics\DIFdelincludegraphics} 
\DeclareRobustCommand{\DIFdelendFL}{\DIFOaddendFL \let\includegraphics\DIFOincludegraphics} 
\makeatletter 
\let\sout@orig\sout 
\renewcommand{\sout}[1]{\ifmmode\text{\sout@orig{\ensuremath{#1}}}\else\sout@orig{#1}\fi} 
\makeatother 
\RequirePackage{listings} 
\RequirePackage{color} 
\lstdefinelanguage{DIFcode}{ 
  moredelim=[il][\color{red}\sout]{\%DIF\ <\ }, 
  moredelim=[il][\color{blue}\uwave]{\%DIF\ >\ } 
} 
\lstdefinestyle{DIFverbatimstyle}{ 
	language=DIFcode, 
	basicstyle=\ttfamily, 
	columns=fullflexible, 
	keepspaces=true 
} 
\lstnewenvironment{DIFverbatim}{\lstset{style=DIFverbatimstyle}}{} 
\lstnewenvironment{DIFverbatim*}{\lstset{style=DIFverbatimstyle,showspaces=true}}{} 
\begin{document}

\title{GHz control of THz QCL band structure and gain by standing acoustic strain}

\author{Alexander S. Kuznetsov}
\email[corresponding author: ]{kuznetsov@pdi-berlin.de}
\affiliation{Paul-Drude-Institut f{\"u}r Festk{\"o}rperelektronik, Leibniz-Institut im Forschungsverbund Berlin e.V., Hausvogteiplatz 5-7, 10117 Berlin, Germany}

\author{Valentino Pistore}
\affiliation{Paul-Drude-Institut f{\"u}r Festk{\"o}rperelektronik, Leibniz-Institut im Forschungsverbund Berlin e.V., Hausvogteiplatz 5-7, 10117 Berlin, Germany}

\author{Lutz Schrottke}
\affiliation{Paul-Drude-Institut f{\"u}r Festk{\"o}rperelektronik, Leibniz-Institut im Forschungsverbund Berlin e.V., Hausvogteiplatz 5-7, 10117 Berlin, Germany}

\author{Klaus Biermann}
\affiliation{Paul-Drude-Institut f{\"u}r Festk{\"o}rperelektronik, Leibniz-Institut im Forschungsverbund Berlin e.V., Hausvogteiplatz 5-7, 10117 Berlin, Germany}

\author{Xiang Lü}
\affiliation{Paul-Drude-Institut f{\"u}r Festk{\"o}rperelektronik, Leibniz-Institut im Forschungsverbund Berlin e.V., Hausvogteiplatz 5-7, 10117 Berlin, Germany}

\date{\today}
\begin{abstract}

Active frequency comb generation and waveform control are central challenges in the terahertz (THz) domain. In THz quantum cascade lasers (QCLs), these functions have typically been achieved through active bias modulation, which alters the operating point of the device and imposes severe limitations on its flexibility. To address these challenges, we propose an approach based on the direct modulation of the QCL bandstructure using GHz-frequency standing bulk acoustic waves (BAWs), promising direct and localized control of the optical gain and chromatic dispersion.
To this end, we fabricated a bulk acoustic transducer on top of a THz QCL in order to excite GHz standing BAWs within its active region. We demonstrate that radio-frequency driving of the transducer leads to the tunable generation of standing BAWs in 5--12~GHz frequency range with wavelengths commensurate to the QCL period length. The effect of the BAW on the QCL bandstructure is revealed by measuring photoluminescence (PL) of the active region, where the BAW strain leads to a considerable modulation of the PL energy up to a few $meV$ around its non-modulated value.
We also develop a model and perform bandstructure simulations to predict the effect of the BAW on the QCL subband structure and gain. These results mark the first demonstration of dynamic bandstructure modulation in a THz QCL using GHz acoustic strain, introducing a fundamentally new paradigm that establishes a powerful synergy between QCLs and BAWs towards coherent control and frequency comb engineering in the THz domain.

\end{abstract}

\pacs{}
\maketitle 




\section{Introduction}
\label{Introduction}

Terahertz quantum-cascade lasers (THz QCLs)~\cite{Faist1994} are essential devices for generating tunable, highly brilliant, and coherent radiation in the so-called THz gap, spanning approximately the 100 GHz to 10 THz range. THz QCLs are crucial for a wide range of applications, including sensing, high-resolution spectroscopy, and metrology, to name a few~\cite{Gao2023}. Despite more than 20 years of consistent progress and refinement of the THz QCL technology, these devices still present considerable challenges in terms of design and control of their emission properties, limiting their flexibility and hindering their widespread application~\cite{Vitiello2015, Lu2024}.

The active region of a THz QCL comprises hundreds of nanometer-scale semiconductor layers that form a sequence of quantum wells (QWs). The QW thicknesses define the subband energies and thus primarily determine the designed emission frequency. Under an applied bias, these subbands align to establish the resonant tunnelling and scattering pathways that enable cascade transport and stimulated THz emission. Importantly, the lasing transition cannot be tuned independently: maintaining efficient operation requires simultaneously preserving multiple resonant alignments throughout the period~\cite{Kumar2009}. As a result, the achievable spectral tunability is constrained not only by the engineered subband spacing, but also by the need to maintain the overall bandstructure alignment that enables the cascade to function~\cite{Lu2024}. 

\begin{figure*}[!htb]
	\centering
		\includegraphics[width=0.9\textwidth, keepaspectratio=true]{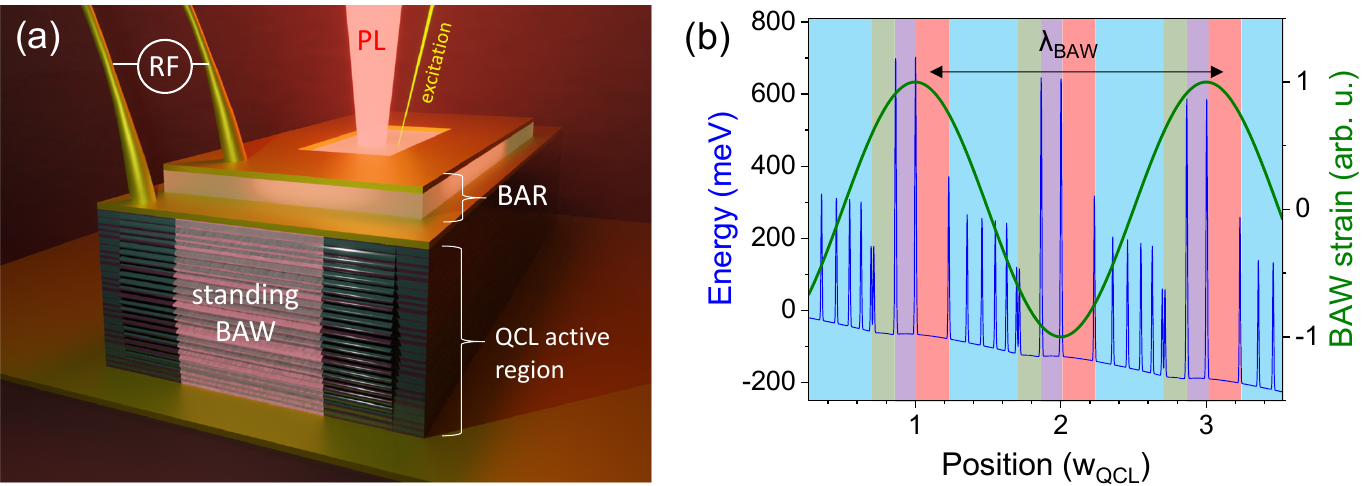}
		\caption{        
            {\bf Structure of the investigated device.}
            {\bf a} Sketch of a quantum cascade laser (QCL) active region modulated by a standing bulk acoustic wave (BAW) generated using a radio-frequency (RF) driven bulk acoustic resonator (BAR). The active region is excited using a non-resonant laser through a small aperture in the BAR and the photoluminescence (PL) is collected and analyzed. 
            {\bf b} Conduction band edge profile of the QCL layer structure over three QCL periods ($w_\mathrm{QCL}$), including interface grading. The color-shaded regions identify the longitudinal-optical phonon transition (green), injector well (purple), lasing well (red) and electron transfer miniband (blue). The strain of a standing BAW with wavelength $\lambda_\mathrm{BAW}$ is indicated with the green curve. 
            }
	\label{Fig1}
\end{figure*}

In this context, one major challenge is the efficient generation of QCL frequency combs, characterized by spectra consisting of highly coherent, equidistantly spaced lines~\cite{Burghoff2014,Silvestri2023}. QCL combs offer a compact architecture together with high output power and spectral tunability across the mid-infrared and THz ranges. Recent advances have enabled QCL combs through intracavity four-wave mixing, dispersion engineering, and bias modulation~\cite{Hugi2012,Wang2017,Heckelmann2023}. However, dynamic control of THz QCL emission remains fundamentally constrained by electrical bias modulation, which perturbs the entire active region uniformly and disrupts the device operation point~\cite{Gellie2010, Wienold2014, Heckelmann2023}. This intrinsic coupling between optical modulation and lasing bias limits the achievable flexibility, bandwidth, and stability of the coherent control.


A different paradigm can be pursued by directly and rapidly modulating the QCL bandstructure to achieve dynamic control over gain and loss. This approach offers a promising route beyond purely electrical modulation, enabling active frequency-comb generation and programmable waveform control of THz emission. It builds on the well-established concept of using MHz and GHz acoustic waves (phonons) to modulate photonic structures~\cite{DeLima2005}. When dynamic acoustic strain is applied to QWs, it modulates the bandgap energy via the deformation potential mechanism~\cite{Akimov2006}. The modulation splits between the valence- and conduction-band edges, thereby directly perturbing the electronic subband structure of the quantum wells forming the QCL active region. As a consequence, the confined energy levels, tunneling couplings, and wavefunction overlaps are dynamically reshaped by the strain. In a QCL, where optical gain arises from intersubband transitions between engineered conduction-band states, even small strain-induced shifts of a few meV can significantly modify level alignment, injection efficiency, population inversion, and oscillator strength. Unlike interband devices, where acoustic modulation primarily affects transition energies, in intersubband systems it directly impacts transport and gain formation.

Effects of surface acoustic wave (SAWs) with sub-GHz frequency on the QCL gain have been theoretically discussed already a decade ago~\cite{Meng2013, Cooper2013}. However, shallow SAW penetration depth and limited frequency range significantly restrict their efficiency for the QCL modulation.
%
%
Conversely, it has been shown that even a single GHz-frequency pulse of bulk acoustic strain, propagating through the QCL active region, leads to a few per cent modulation of the THz emitted power~\cite{Dunn2020}. However, such broadband, ultrafast strain excitations perturb only a small portion of the active region and rely on external picosecond optical pumping, constraining both scalability and performance. Moreover, no QCL model has yet been developed that can incorporate ultrafast dynamical strain effects on the gain and dispersion, limiting the investigation to a purely experimental effort.

Here, we aim at a fundamentally different level of control by employing GHz-frequency standing bulk acoustic waves (BAWs) generated along the QCL growth axis to coherently modulate its bandstructure. By integrating piezoelectric transducers with frequencies tunable in the 5-12 GHz range directly onto the QCL top contact, we generate strong, standing strain fields.
Photoluminescence measurements of the emission from the QCL active region confirm modulation of the GaAs bandgap and active region energies by GHz BAWs. 
Our theoretical model shows that these experimentally demonstrated modulation amplitudes can significantly influence the QCL gain and dispersion, providing the degrees of freedom required for coherent control of THz QCL emission without affecting the operating point of the device.




\section{Results}
\label{Results}

Figure~\ref{Fig1}a shows a schematic of our implementation of a THz QCL with an integrated bulk acoustic wave resonator (BAR) fabricated on top of the active region. Details of the actual sample and fabrication procedure are discussed in Methods. First, QCL layers are grown by molecular beam epitaxy on a double-side polished GaAs substrate~\cite{Lu2024}. Then, the QCL waveguide is wet-etched into a $120~\mu$m-wide and $10.8~\mu$m-thick mesa, followed by the BAR fabrication. The BAR generates BAW when driven with a radio-frequency signal (RF) which can be delivered via RF probes or wire bonding. The generated BAW creates a standing wave pattern within the QCL active region through multiple reflections between the top QCL surface and the substrate backside~\cite{Machado2019}. In order to monitor photoluminescence (PL) from the active region under acoustic modulation, a BAR has $\mu$m-sized openings in both the top and bottom metallic contacts, the latter being also the top contact of the QCL waveguide. An excitation laser can hence be focused on the aperture to excite the QCL active region PL, which is then collected and analyzed using a spectrometer, as described in Methods. 


\begin{figure*}[!htb]
	\centering
		\includegraphics[width=0.9\textwidth, keepaspectratio=true]{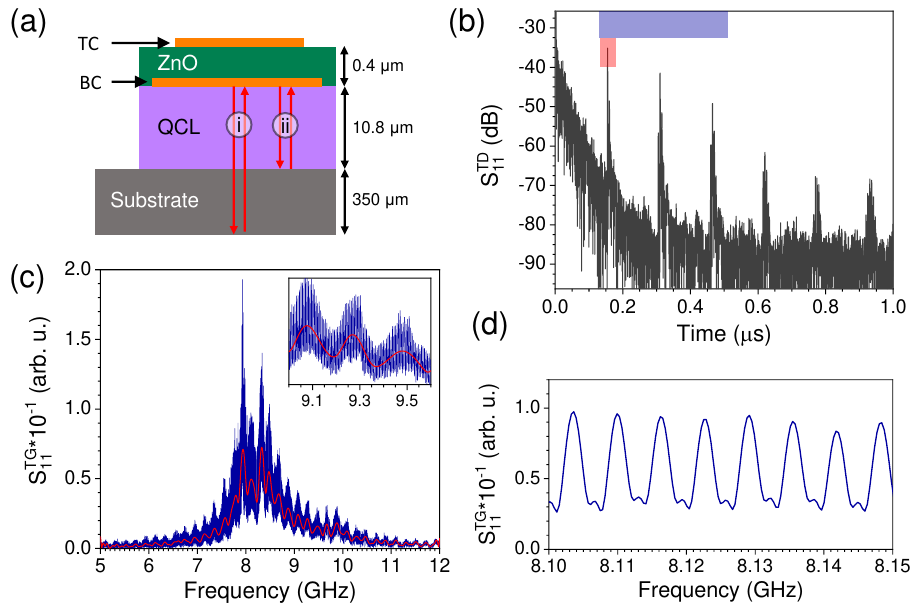}
		\caption{
            {\bf BAR response on a QCL.}
            {\bf a} Schematic side-view of a BAR on a QCL (not-to-scale). The BAR consists of a top (TC) and a bottom (BC) metal contacts and a thin film of piezoelectric ZnO. The active region of the BAR is defined by the overlap between TC and BC. The vertical red arrows indicate two possible propagation paths for the acoustic wave: (i) through the whole structure, and (ii) localized in the QCL.
            {\bf b} Measured time-domain $s_{11}^{\text{TD}}$ RF reflection coefficient of a BAR on a QCL. The red and blue rectangles depict time-gate regions.
            {\bf c} Time-gated $s_{11}^{\text{TG}}$ curves of the first BAW (red line) and first three (dark blue line) echoes in {\bf b} indicated by the red and blue rectangular areas, respectively. The inset shows a close-up of low-periodicity oscillations.
            {\bf d} A zoom of the high-periodicity oscillations in $s_{11}^{\text{TG}}$ in {\bf c}. 
            }
	\label{Fig2}
\end{figure*}

Figure~\ref{Fig1}b illustrates the principle of QCL bandstructure modulation using BAWs. It shows one possible spatial alignment of the standing BAW strain and the QCL bandstructure across several periods of the biased QCL active region, which in this study is based on a hybrid design with the lasing frequency at about 4.7 THz~\cite{Kohler2004, Scalari2005, Wienold2009}. The general idea is to employ a BAW with a spatial wavelength ($\lambda_\text{BAW}$) commensurate with the QCL period length ($w_\text{QCL}$), i.e., to achieve a condition where the ratio $\lambda_\text{BAW} / w_\text{QCL}$ is a rational number. In this way, the spatial positions of the BAW nodes and antinodes remain fixed across all active-region periods, enabling targeted modulation of specific regions within each period.


Practically, for the QCL used in this work, $w_\text{QCL} = 122.84$~nm. Thus, in order to realize the condition displayed in Fig.~\ref{Fig1}b, the BAW wavelength must satisfy $\lambda_\text{BAW} = 2w_\text{QCL} = 245.68$~nm. For a GaAs-dominated QCL structure grown along the [100] direction, this condition corresponds to an acoustic frequency of $f_\mathrm{BAW} = v_\text{BAW} / \lambda_\text{BAW} = 19.1$~GHz, where $v_\text{BAW} \approx 4.7 ~ \mu$m/ns is the longitudinal BAW velocity. Previously, we have demonstrated BARs operating at 20~GHz~\cite{Machado2019,Kuznetsov2021}. However, such high-frequencies impose stringent constraints on fabrication and device geometry, for example requiring sub-100-nm-thick piezoelectric ZnO layers. For this proof-of-principle study, we therefore employ BARs operating in the 7--10~GHz frequency range, for which the corresponding BAW wavelength spans multiple QCL periods.


A BAR device comprises a simple capacitor-like structure, where the dielectric is replaced with a thin film of a piezolectric layer, as schematically shown in Fig.~\ref{Fig2}a. The BAR was designed to generate around $f_\mathrm{BAR} = 8.3$~GHz, which corresponds to $\lambda_\mathrm{BAW} = v_\mathrm{BAW} / f_\mathrm{BAR} \approx 570~\text{nm} \approx 4.7 w_\text{QCL}$. Consequently, the acoustic standing-wave profile cannot be perfectly matched to the intrinsic periodicity of the active region in this case, and a one-to-one alignment between acoustic antinodes and specific layers within each period is not achieved in the present devices. It is important to stress, however, that the primary objective of this work is to demonstrate the coherent generation of BAWs within a THz QCL active region and to establish, through systematic modeling and simulation, their impact on the bandstructure and optical properties. Precise matching between the BAW wavelength and the active-region period will be considered directly within the theoretical model.

Experimentally, we first consider the generation of GHz BAWs using BARs fabricated on top of the active region. In the following, all measurements were carried out in the 5-10 K temperature range and with no bias applied to the active region. This does not limit the relevance of the study, since the BAWs couple directly to the lattice through strain. Operating without electrical bias, however, eliminates parasitic effects associated with carrier injection and Joule heating, ensuring that the observed response arises purely from the acoustic modulation of the bandstructure.

Piezoelectric transducers can be used for electrical detection of phonons using a vector network analyzer (VNA)~\cite{Machado2019}. Figure~\ref{Fig2}b shows the time-domain power reflection coefficient $s_{11}$ of a BAR in the 5--12~GHz range recorded at 10~K temperature. Above 100 ns, the signal consists of narrow equidistant peaks, which originate from BAW echoes corresponding to the round-trip time of the BAW through the QCL and the substrate. The blue curve in Fig.~\ref{Fig2}c presents the time-gated frequency dependence of $s_{11}$ corresponding to the first three echoes in Fig.~\ref{Fig2}b. The curve can be understood as the measure of the BAW transmission through the QCL and the substrate. The Gaussian-shaped response peaks at about 8.3~GHz and has a bandwidth of around 7 GHz. A close inspection of the smaller frequency range, cf. Fig.~\ref{Fig2}d, reveals a fine comb of resonances with a free spectral range of 6.3~MHz, which are the modes of the standing BAW in a Fabry–Pérot-like resonator formed by the entire structure, as indicated by the path (i) in Fig.~\ref{Fig2}a. The red curve in Fig.~\ref{Fig2}c shows the single-echo response, which clearly reveals additional modulation of the BAR spectrum. This modulation is due to the active region acting like an acoustic cavity for BAWs due to the mismatch between the effective acoustic impedance of the active region and the one of the GaAs substrate, as indicated by the path (ii) in Fig.~\ref{Fig2}a.

\begin{figure*}[!htb]
	\centering
		\includegraphics[width=0.9\textwidth, keepaspectratio=true]{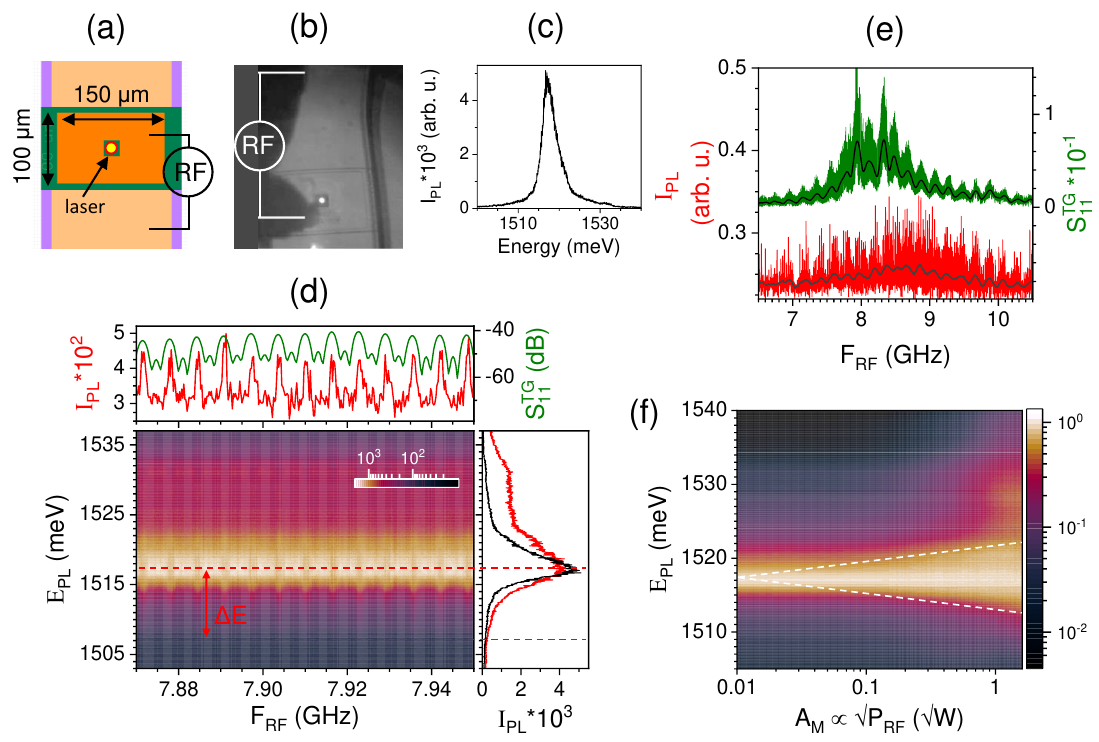}
		\caption{
            {\bf BAW modulation of a QCL active region.}
            {\bf a} Schematic top-view of the QCL waveguide with a BAR with an aperture for PL measurements. 
            {\bf b} A camera image of the actual QCL-BAR device. The RF excitation is provided by the probes (dark shape). The bright spot on the aperture is the focused laser beam. The curving of the lines of the BAR and the waveguide is an artifact.
            {\bf c} PL spectrum of the QCL active region with RF off.
            {\bf d} Central section, PL spectrum as a function of RF frequency ($F_{\text{RF}}$) applied to the BAR for a fixed RF power of $P_{\text{RF}} = 29$~dBm ($0.89 \sqrt{\text{W}}$). The right inset shows two spectra for the out-of-phase condition (black curve) and in-phase one for $F_{\text{RF}} = 7.91$~GHz (red curve) resonance cases. The upper inset compares the PL intensity profile around 1513~meV (red curve) and the time-gated s-parameter (green curve).
            {\bf e} Comparison of the intensity profile at 1513~meV extracted from {\bf d} (red curve) with the time-gated s-parameter (green curve) over the 6.5--10.5~GHz range from Fig.~\ref{Fig2}c. The black curves superimposed on the PL and $s_{11}$ show smoothed-by-averaging PL intensity profile and first echo $s_{11}$, respectively.
            {\bf f} Normalized PL spectrum as a function of the acoustic amplitude ($A_{\text{M}} \propto \sqrt{P_{\text{RF}}}$) applied to the BAR for $F_\text{RF} = 7.91$~GHz.
            }
	\label{Fig3}
\end{figure*}

The effect of the BAW strain on the bandgap of the QCL can be conveniently investigated using PL measurements~\cite{Kuznetsov2021}. Figure~\ref{Fig3}a schematically shows the top-view of a BAR device with a small aperture in the electrical contacts for optical access. In the experiment, we focus a 760~nm continuous wave laser (0.1~mW power) on the aperture and measure PL. A photograph of the sample, cf. Fig.~\ref{Fig3}b, shows RF probes in contact with the BAR and the bright laser spot on the aperture. The PL of the active region is characterized by a broad peak centered at around 1515~meV as shown in Fig.~\ref{Fig3}c. Due to the quantum well-like character of the GaAs layers of the QCL, the PL of the active region is drastically different from that of the GaAs substrate, as demonstrated in Appendix A1.
Due to the large absorption of the QCL active region, which is approximately 97.7\% GaAs (cf. Fig.~\ref{Fig1}b), the excitation penetration depth at 760~nm is $L_{\text{pen.}}^{\text{opt.}}\simeq 0.5 ~\mu$m~\cite{Sturge1962}. This value sets the scale of the optical probing of the QCL-phonon interaction.

Applying high-power RF driving to a transducer can lead to undesirable local heating of the sample, which would introduce additional spectral redshift, not related to the acoustic modulation. To reduce the thermal loading effect, instead of continuous excitation, both RF and laser stimulation were pulsed and synchronized, as described in detail in Appendix A2. In a typical experiment, pulses of $\sim 600 ~ \mu$s width and 10\% duty cycle were used. The RF measurements discussed in the following were performed using this procedure.

The color map of Fig.~\ref{Fig3}d shows the dependence of the PL spectrum on the RF frequency applied to the BAR in the 7.87--7.95~GHz range for a fixed nominal RF power of 29~dBm. Multiple very sharp diamond-shaped resonances separated by about 6.4~MHz can be easily identified. At each resonance, the time-integrated PL spectrum is broadened in energy, as shown in the inset on the right. The red- and blue- shifts in energy are characteristic of the effect of BAW strain on the bandgap~\cite{DeLima2005}. We note that the maximum detected shifts of the PL energy reach at least $\Delta E \geq 10$~meV, as indicated in the Fig.~\ref{Fig3}d by the horizontal dashed lines. The value of the $\Delta E$ is limited by the maximum available power of the RF generator. We note that the exact determination of the energy modulation amplitude is complicated by the large spectral width of the PL and due to the fact that the PL probes depth region corresponding to the full acoustic wavelength, i.e., $L_{\text{pen.}}^{\text{opt.}} \approx L_{\text{BAW}}$. 

The upper inset of Fig.~\ref{Fig3}d compares the PL intensity profile integrated over a 0.1~meV energy range around 1513~meV (red curve) and the time-gated $s_{11}$ (green curve). The comparison shows a perfect match between the resonances in PL and acoustic response of the transducer. The same curves are shown in Fig.~\ref{Fig3}e over the whole 5--12~GHz frequency range. It is clear that the modulation spans the whole generation bandwidth of the BAR. The averaged PL spectrum (black curve) reveals the same large frequency scale modulation as the $s_{11}$ curve, due to the acoustic cavity effect of the QCL active region. 
We also note there is an appreciable relative shift between the PL spectrum and the $s_{11}$ in Fig.~\ref{Fig3}e. This discrepancy can be attributed to the fact that the acoustic modes that most efficiently modulate the PL are not necessarily those that are most efficiently excited by the BAR, as the regions of BAW excitation and PL emission are located at different depths within the structure.

A dependence of the PL spectrum on the acoustic amplitude for a fixed frequency $F_\mathrm{RF} = 7.9$~GHz is shown Fig.~\ref{Fig3}f. The spectrum broadens symmetrically with increasing BAW amplitude, which is characteristic of the deformation potential modulation mechanism. For the largest BAW amplitudes, the energy modulation amplitude reaches several meV. Owing to the pulsed nature of the experiment, the RF-induced local heating of the sample due to the imperfect RF-to-BAW conversion by the BAR is negligible, as is demonstrated in Appendix A3 by comparing in-phase and out-of-phase measurements. 

\begin{figure*}[!t]
	\centering
		\includegraphics[width=0.9\textwidth, keepaspectratio=true]{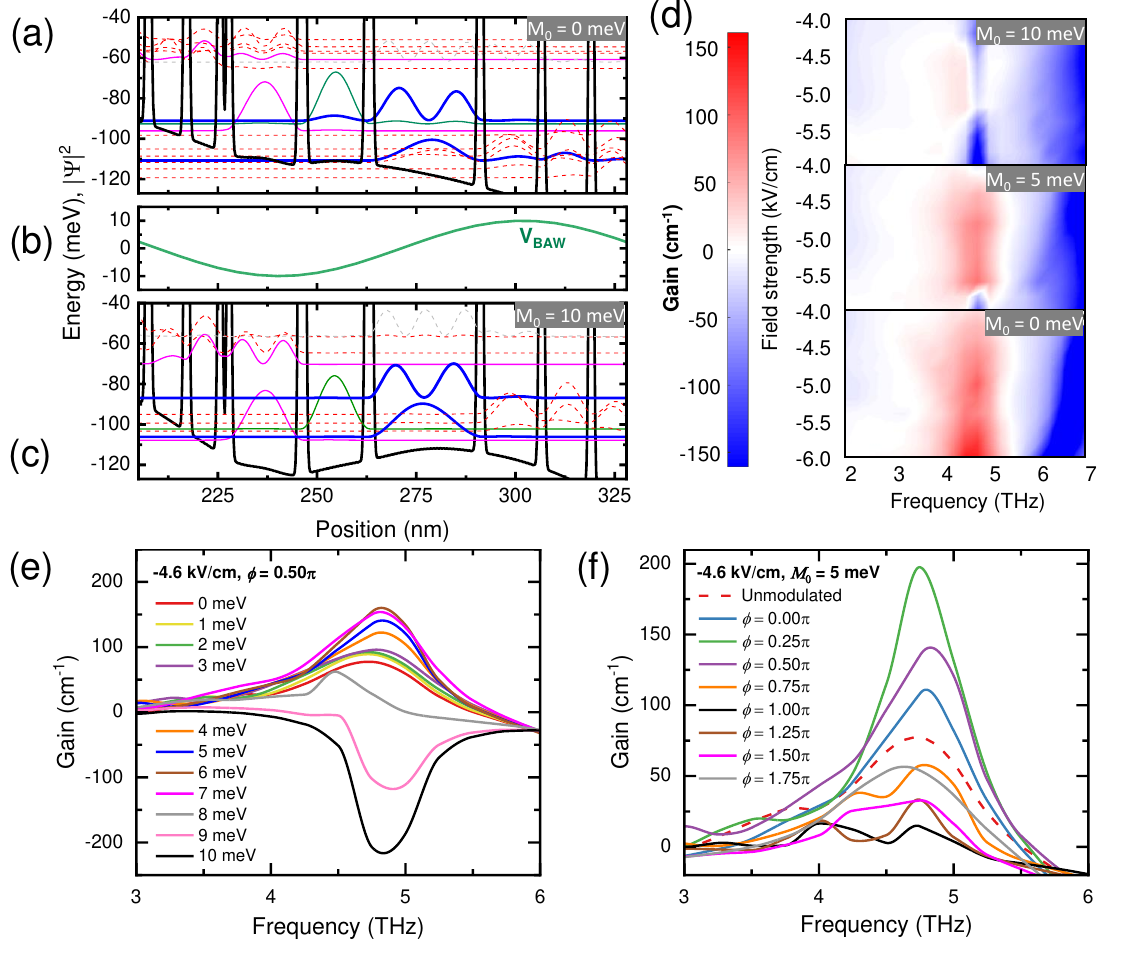}
		\caption{
            {\bf Theory: QCL gain under the BAW modulation.}
            {\bf a} Simulated potential profile and subband structure (moduli squared of the envelope functions) for a 4.75 THz QCL active region at a field strength of -4.6 kV/cm and without any BAW modulation. The lasing levels are in blue, the injection level in green, and the LO-phonon transition levels are in violet. The levels of the miniband are in dashed red.
            {\bf b} Spatial profile assumed for the BAW potential ($V_\text{BAW}$) with the amplitude $M_{0} = 10$~meV. The spatial phase ($\phi = \pi$) has been set to show a maximum at the extractor barrier, which leads to a strong energy mismatch between the injector level and the upper laser level.
            {\bf c} The same simulated subband structure as in (a), with a applied BAW potential in (b). 
            {\bf d} Gain maps at three BAW amplitudes and $\phi$ = 0. One can notice the switch from gain to losses for all field strengths at higher BAW modulations.
            {\bf e} Exemplary gain profiles for -4.6~kV/cm bias for BAW amplitudes from 0~meV to 10~meV and $\phi$ = $0.50 \pi$. 
            {\bf f} Exemplary gain profiles for -4.6~kV/cm bias for various BAW phases and a modulation amplitude $M_\mathrm{0}$ = 5 meV.
            }
	\label{Fig4}
\end{figure*}

For the simulation of THz QCLs bandstructure, theoretical models with varying levels of complexity and computational cost have been realized~\cite{Jirauschek2014}. For practical device design, however, a scattering-rate approach based on the self-consistent solution of the Schrödinger and Poisson equations proves particularly effective. In this method, the scattering rates are iteratively calculated together with the carrier distribution and Coulomb potential. The corrected potential is then reinserted into the Schrödinger–Poisson system, and the process is repeated until convergence is achieved~\cite{Schrottke2010}.

To efficiently capture the relevant transport and optical properties, we developed a phenomenological model in which all quantities are formulated in Fourier space~\cite{Schrottke2015,Lu2016}. This framework enables the computation of full current density–field strength characteristics and the corresponding gain maps within only a few minutes. For a BAW frequency of approximately 10 GHz, corresponding to a period of about 100 ps, the system operates in a quasi-adiabatic regime for the electrons. In THz QCLs, electronic scattering processes and wavefunction relaxation occur on timescales of a few picoseconds~\cite{Dunn2020,Paiella2001,Bacon2016}, which are much shorter than the acoustic period. As a result, the energy modulation induced by the BAW can be incorporated as a static correction to the potential, assuming that the electronic populations respond instantaneously to the slowly varying acoustic potential. As discussed in more detail in the Discussion section, this approximation is well justified for the intersubband carrier dynamics characteristic of THz QCLs.

Within our Fourier-transform-based model, the acoustic perturbation is introduced as an extra potential term $V(z) = M_{0} \cos(2 \pi z / \lambda_\text{BAW} + \phi)$ where $M$$_0$ is the modulation amplitude, $\phi$ the acoustic phase, and the simulation cell length is chosen as $w_\text{QCL} = \lambda_\text{BAW}$. This choice is dictated by the periodic boundary conditions employed in the model. Simulating a sub-harmonic of the BAW, i.e., $w_\text{QCL} = \lambda_\text{BAW}/N$ with $N$ a natural number, would introduce artificial discontinuities at the boundaries. Conversely, considering a superperiod comprising multiple QCL periods leads to non-uniform local electric fields across the structure, a situation referred to as quasi field-domain formation. This effect is further enhanced in hybrid active-region designs, such as the one considered here, due to the strong local charge accumulation associated with the injector-induced dipole. Nevertheless, an approximate treatment of these scenarios is possible and is discussed in Appendix A4.

We assume a static splitting of the modulation, with 20 percent acting on the valence band and 80 percent on the conduction band, due to the fact that the conduction states are primarily s-like and are more sensitive to the hydrostatic strain-induced changes than the p-like valence states~\cite{Warburton1991,DeLima2005}. Within this framework, the temporal evolution of the optical gain can then be reconstructed by evaluating the active-region response for different modulation amplitudes, each corresponding to a specific time within the BAW cycle. Predicting the resulting THz waveform emitted by the laser, however, is considerably more challenging even in the absence of acoustic modulation, as illustrated by the most advanced models developed to tackle this task~\cite{10.1016/j.cpc.2021.108097, 10.1364/OE.24.023232}. For our devices, such simulation framework would also require a full time-domain treatment of the coupled carrier–photon dynamics. In the Discussion section we therefore analyze a representative simplified case to provide qualitative insight, while a comprehensive modeling for the general problem will be the subject of future work.

Figure ~\ref{Fig4}a shows the simulated potential profile and the squared moduli of the wave functions of a single period of the 4.7 THz hybrid-design active region employed in the PL experiments, at a field strength of -4.6 kV/cm, in the absence of BAW modulation. The lasing states, injector level, longitudinal-optical (LO) phonon transition levels, and miniband states are highlighted in different colors for better visibility. At this operating field, the injector level is well aligned with the upper laser level, enabling efficient injection, population inversion, and optical gain. For reference, a QCL cleaved from the same wafer was characterized electrically and optically. The corresponding light output-current-voltage ($L$-$I$-$V$) curve and emission spectrum are reported in Appendix A5.

The effect of the acoustic modulation, with the profile shown in Fig.~\ref{Fig4}b, on the bandstructure is illustrated in Fig.~\ref{Fig4}c, where the same period is subjected to a BAW with amplitude $M$$_0$ = 10 meV. The phase $\phi$ is chosen such that a modulation maximum aligns with the extractor barrier and a minimum with the barrier between the LO-phonon transition well and the injector well. In this configuration, the relative alignment of the relevant states is strongly altered: the lasing levels shift upward in energy, while the injector level shifts downward. As a result, carrier injection preferentially occurs into the lower laser level, suppressing population inversion and leading to pronounced optical absorption instead of gain. This example highlights the strong sensitivity of the QCL bandstructure and therefore of the optical response to the spatial profile of the acoustic strain.

A systematic view of this behavior is provided in Fig.~\ref{Fig4}d, which reports simulated gain maps for different values of $M_0$ at fixed phase $\phi$ = 0, corresponding to modulation nodes aligned with the lasing transition. Here one can notice how the BAW affects the gain differently for different values of the field strength. We then aim to decouple the effect of the field strength from that of the BAW parameters. To do so, we show in Fig.~\ref{Fig4}e the gain spectra at -4.6 kV/cm and $\phi$ = 0.50$\pi$ for $M_0$ ranging from 0 to 10 meV. As the modulation amplitude increases up to about 7 meV, the peak gain rises. For larger amplitudes, however, the gain rapidly decreases and turns into strong net absorption at 9–10 meV. This non-monotonic behavior reflects the delicate balance between injection, depopulation, and level alignment under strain. In this regard, the acoustic phase provides an additional and powerful control parameter. In Fig.~\ref{Fig4}f, gain spectra are shown at a fixed field and $M$$_0$ = 5 meV while varying $\phi$ from 0 to 1.75$\pi$. Depending on the phase, the gain can be enhanced, reduced, or nearly flattened across the emission bandwidth, demonstrating that phase-controlled strain enables fine tuning of both the amplitude and spectral shape of the gain. 

The gain enhancement can be understood by noting that the system is simulated near the center of its operating range, where an increase in the effective electric field leads to a corresponding increase in gain. In this regime, the BAW-induced strain can play a similar role by locally modifying the bandstructure and improving the alignment between the injector states and the upper lasing level, thereby enhancing carrier injection into the lasing transition.

Overall, the introduction of BAWs substantially modifies the electronic wavefunctions and the relative alignment of the subbands, resulting in pronounced changes in both the magnitude and spectral distribution of the optical gain. The strong dependence on acoustic amplitude and phase demonstrates that localized bandstructure modulation provides a versatile and efficient means to engineer the QCL gain and dispersion through acousto-electronic control.

Finally, we note that the bandstructure simulations describe the steady-state response of the system, providing only instantaneous “snapshots” of the gain at specific instants of the modulation. The effective dynamical response of the structure will be discussed in the following section.

\section{Discussion}
\label{Discussions}

Firstly, we discuss the experimental findings. We stress that the amplitude of the strain-induced energy modulation demonstrated here represents a lower limit of the actual modulation. Indeed, we measure PL away from the active region of a BAR. The latter excites predominantly a longitudinal BAW that propagates perpendicularly to the surface. There is, however, a small diverging component that propagates laterally, as explained in Appendix A6. Hence, the observed modulation originates from the interference of multiple acoustic reflections responsible for the lateral propagation of the acoustic wave. While we do not have a direct measure of the modulation amplitude beneath the opaque BAR electrodes, we expect considerably larger strain amplitudes directly below the BAR active region.

We emphasize that the experimentally demonstrated modulation of the QCL bandstructure is a prerequisite for achieving full dynamical control of the THz emission. In the present case, the maximum frequency with non-negligible energy modulation was limited to $\approx 10$~GHz, corresponding to the acoustic wavelength $\lambda_{\text{BAW}} \approx 470~\text{nm} \approx 4 w_\text{QCL}$. Precise matching between the BAW wavelength and the active-region period will be addressed experimentally in future studies on fully operational QCL devices integrating optimized BAR designs.


To reach the condition $\lambda_{\text{BAW}} = 2 w_{\text{QCL}}$ corresponding to the one depicted in Fig.~\ref{Fig1}b, the frequency would need to be increased to approximately 20~GHz for this active region design. Efficient piezoelectric excitation of 20~GHz BAWs and subsequent acoustic modulation of QW energies has already been demonstrated using 80 nm thick piezoelectric ZnO films~\cite{Kuznetsov2021}. In order to have $\lambda_{\text{BAW}} = w_{\text{QCL}}$, a BAW frequency of 40~GHz will be required. Recent progress in the fabrication of high-overtone BARs operating up to 120~GHz~\cite{Kramer2025} makes this a realistic possibility. However, generating BAWs at such high frequencies generally comes at the cost of reduced efficiency and smaller strain amplitudes. This limitation could be mitigated by fabricating QCLs on a phononic distributed Bragg reflector (DBR), which would concentrate the strain field at the desired frequency within the active region. Alternatively, a longer active-region period can be employed, which is typically associated with a lower lasing energy for the same design. For instance, for an already demonstrated 3.5~THz active region~\cite{Lu2024}, the longer period reduces the required frequency to about 28~GHz.

The experiments of this work were carried out at 10~K. For practical applications, operation above 50~K may be required, where Stirling cryocoolers can be used. The performance of the BAR itself does not strongly depend on temperature and can in principle operate even at room temperature. However, the absorption of acoustic phonons in GaAs increases by approximately a factor of four between 10~K and 100~K, leading to a reduction of the standing-wave ratio (SWR). To mitigate this effect, the aforementioned acoustic DBR could be fabricated beneath the QCL in order to shorten the acoustic cavity and increase the SWR.

Our simulation framework predicts the instantaneous response of the QCL bandstructure and the corresponding optical gain under BAW modulation by assuming that the characteristic timescales of the relevant intersubband processes are much shorter than the acoustic period. Under this condition, the acoustic field can be considered effectively frozen in time from the perspective of the electronic dynamics.

This assumption is well justified by the large mismatch between the relevant timescales. For acoustic frequencies below 10~GHz, the modulation period ($T_{\text{BAW}}$) exceeds 100~ps. In contrast, the fundamental microscopic processes governing carrier transport in THz QCLs occur much faster: LO-phonon scattering takes place on a timescale of $\approx$1~ps, injector tunneling in $\approx$2~ps, and intersubband dephasing in well below 1~ps (Table~\ref{table1}). Even photon lifetimes in the cavity are on the order of $\approx$10~ps, still significantly shorter than the acoustic modulation period considered here. As a result, electrons and photons effectively follow the slowly varying acoustic potential quasi-instantaneously, validating the quasi-adiabatic approximation adopted in our model.

Predicting the emitted THz waveform of a BAW-modulated QCL is considerably more challenging. The laser output depends on many device- and system-specific parameters, including the waveguide type (e.g., single-plasmon or double-metal), the resonator geometry (Fabry–Perot, ring, surface-emitting, random, etc.), the fraction of the cavity volume modulated by the acoustic field, the number and spatial distribution of BARs, their driving conditions, the cavity length and associated photon round-trip time ($\tau_{\text{rt}}$), the active-region design, and the operating bias of the QCL. Accurately predicting the time-dependent emission of actively modulated THz QCLs is itself a challenging problem that has only recently been addressed, for example in studies of active mode locking~\cite{Schreiber2025}. Extending such time-domain simulation frameworks to the case of BAW-driven modulation would represent a substantial effort beyond the scope of the present work and will require a dedicated study. At the same time, the large number of available control parameters, particularly when multiple BARs are integrated along the same waveguide, suggests that acoustic modulation could enable levels of control over the THz output well beyond conventional bias-based active mode locking.

Within this broader context, it is nevertheless instructive to consider a simplified configuration that can provide qualitative insight into the expected laser dynamics. As an illustrative example, one may consider a THz QCL whose active region is uniformly modulated by a single BAR. The RF driving of the BAR is assumed to be spatially uniform along the device, while the QCL operates well above threshold and sufficiently far from the negative differential resistance (NDR) region. In addition, we assume that the BAW wavelength is matched to the active-region periodicity, with fixed modulation amplitude and phase.

Under these conditions, three different regimes can be identified depending on the relation between $\tau_{\text{rt}}$ and $T_{\text{BAW}}$. When $\tau_{\text{rt}} \gg T_{\text{BAW}}$, photons circulating in the cavity experience a time-averaged gain determined by the instantaneous gain values corresponding to different acoustic modulation amplitudes (Fig.~\ref{Fig4}e). For small modulation amplitudes and neglecting nonlinearities, the laser therefore operates close to this average gain value, as the intrinsically fast gain dynamics of the QCL stabilizes the intracavity field around its mean level~\cite{Khurgin2014}. If the acoustic modulation significantly perturbs the gain, for example periodically driving it below threshold, the modulation can imprint onto the emitted waveform, potentially producing nonlinear dynamics similar to those observed in typical active mode locking. Such behavior may be further enhanced when operating closer to threshold or near the onset of the NDR region, where small perturbations of the subband alignment produce large variations in the net gain.

In practice, however, achieving the condition $\tau_{\text{rt}} \gg T_{\text{BAW}}$ at acoustic frequencies around $F_{\text{BAW}} \sim 10$~GHz would require cavity lengths in the order of $\sim 1$~cm or longer, far beyond the typical dimensions of THz QCLs, which rarely~\cite{Wienold2014} exceed 4~mm.

A second regime corresponds to $\tau_{\text{rt}} \ll T_{\text{BAW}}$, achievable with very short cavities (typically shorter than 1~mm). In this case, each photon round trip samples essentially the instantaneous gain, so that the emission envelope adiabatically follows the acoustic modulation. In principle, the modulation speed is limited by the photon lifetime in the cavity (about 10~ps). In practice, however, shortening the cavity increases mirror losses inversely with cavity length, raising the threshold and limiting device performance.

A third regime arises when $\tau_{\text{rt}} \approx T_{\text{BAW}}$. Successive round trips then probe different phases of the acoustic modulation, resulting in different effective gains at each pass. If the ratio $\tau_{\text{rt}}/T_{\text{BAW}}$ is rational, the modulation and cavity phases periodically realign and the emission becomes periodic. If the ratio is irrational, quasi-periodic dynamics arise. In either case, strong amplitude modulation can occur, producing burst-like emission patterns or periodic pulse trains when $\tau_{\text{rt}} = T_{\text{BAW}}$, similar to conventional active mode locking. In the frequency domain, cavity modes acquire sidebands spaced by the acoustic modulation frequency.

From this perspective, the above configuration can be viewed as the acoustic analogue of traditional active mode locking implemented through bias modulation. The key difference lies in the physical mechanism used to modulate the gain: instead of electrical bias modulation, the present approach directly perturbs the QCL bandstructure through coherent acoustic strain.

Table~\ref{table1} also shows that not all relevant dynamical processes in THz QCLs occur on ultrafast timescales. In particular, the gain recovery time (15–50 ps) approaches the period corresponding to modulation frequencies in the tens-of-GHz range. Recent demonstrations of BAR devices operating at such frequencies therefore indicate that acoustic modulation can occur on timescales comparable to the intrinsic gain dynamics of the laser. In this regime, the acoustic perturbation can no longer be regarded as quasi-static with respect to the carrier populations and intracavity field, opening access to a genuinely non-adiabatic operating regime.

This perspective is particularly intriguing in the context of Floquet engineering and synthetic frequency-domain dynamics. While the fastest microscopic processes (e.g., LO-phonon scattering or dephasing) are unlikely to be directly driven coherently due to their sub-picosecond timescales, collective quantities such as population inversion, modal gain, and intracavity field evolution occur on tens-of-picoseconds timescales that are experimentally accessible with high-frequency BAWs. Periodic strain at these frequencies can therefore drive the system into a genuinely time-dependent regime, dynamically modulating gain recovery and mode competition processes, and enabling controlled reshaping of the laser spectrum. At the same time, such periodic driving naturally gives rise to coherent coupling between longitudinal cavity modes separated by the modulation frequency, effectively forming a synthetic frequency lattice~\cite{Heckelmann2023}. In this picture, the comb modes correspond to lattice sites, while the acoustic modulation provides the coupling between them. The fast gain recovery characteristic of THz QCLs further introduces nonlinear interactions, such as dissipative four-wave mixing, which can drive synchronization and phase locking across the synthetic lattice. This mechanism enables the emergence of frequency-comb states and opens the possibility of realizing more complex dynamics, including nonlinear quantum walk behavior in the frequency domain.

Taken together, these effects establish a unified framework in which coherent acoustic fields act as a programmable resource for both time-domain (Floquet) and frequency-domain (synthetic lattice) control. This hybrid opto-acoustic platform therefore provides a powerful route toward frequency-comb formation, waveform shaping, and the exploration of acousto-optically driven quantum photonic phenomena in THz QCLs.

\begin{table}
  \centering
\begin{threeparttable}[htbp]
  \caption{Temporal characteristics of the BAW and QCL.}
  \begin{tabular}{| c | c  | c |}
 \hline
 \hline
 Parameter   &  Value  &   Comment \\
 \hline
 $T_{\text{BAW}}$     &   118 (50) ps     &   BAW period at 8.5~GHz (20~GHz)  \\ 
 $\tau_{\text{LO}}$     &   $\approx 1$~ps    &   LO-phonon scattering time~\cite{Becker2002}  \\ 
 $\tau_{\text{inj}}$     &   $\approx 2$~ps    &   Injector tunneling time~\cite{Jirauschek2014}  \\ 
 $\tau_{\text{deph}}$     &   $\approx 0.3$~ps    &   Intersubband dephasing time~\cite{Callebaut2005} \\ 
 $\tau_{\text{gain}}$     &   $15 - 50$~ps    &   Gain recovery time~\cite{Bacon2016,Derntl2018,Green2009}  \\ 
 $\tau_{\text{phot}}$     &   $\approx 10$~ps    &   Photon lifetime~\cite{Agnew2015} \\ 
 $\tau_{\text{rt}}$     &   $25 - 75$~ps    &   Cavity photon round-trip time\tnote{1}  \\ 
\hline    
\hline
\end{tabular}
  \begin{tablenotes}
    \item[1] For 1–3 mm cavity and standard waveguide geometry (refractive index $n \approx 3.6$  for GaAs at THz frequencies).
  \end{tablenotes}
\label{table1}
\end{threeparttable}
\end{table}

Although the intrinsic modulation bandwidth of THz QCLs has been estimated to exceed 100~GHz, achieving such rates through bias modulation alone is highly challenging. Even in double-metal waveguides, aggressive RF optimization of the electrical response inevitably compromises optical confinement and guiding of the THz mode. In contrast, acoustic modulation directly perturbs the bandstructure rather than the electrical bias, effectively decoupling the high-frequency drive from the electrical injection pathway. This separation removes a central bottleneck of purely electrical schemes and may constitute the most practical route toward a robust Floquet-engineered regime in THz QCLs. Importantly, the approach is largely independent of the THz waveguiding platform and remains compatible with single-plasmon waveguides that offer higher output power and lower beam divergence.

A further advantage of BAW-based modulation is its intrinsically local and spatially selective character. In contrast to current modulation, which perturbs the bias field uniformly across all cascade periods, acoustic strain couples locally to the electronic states, with an efficiency determined by the spatial overlap between the acoustic field and the relevant wavefunctions. Specifically, the phase (positions of the nodes and anti-nodes) of the BAW can be fine-tuned independently through the acoustic frequency.
%
%
Varying the acoustic phase shifts the standing-wave pattern along the growth axis, thereby controlling the spatial position of the modulated region within each period. Such spatially resolved control enables modulation schemes that are inaccessible with global current injection, including distributed or phase-shifted patterns along the cavity and the engineering of tailored spatio-temporal gain landscapes. In particular, the integration of multiple independently driven BARs along the waveguide naturally leads to a spatially structured and dynamically reconfigurable modulation of the gain and dispersion. From this perspective, the system can be viewed as an active, time-dependent photonic structure, in which the acoustic field defines a programmable lattice for the optical modes. Unlike conventional photonic crystals, where periodicity is fixed by fabrication, the present approach enables dynamic control of both the spatial profile and temporal evolution of the modulation. This opens the possibility of realizing reconfigurable gain–loss gratings and, more generally, spatio-temporally modulated photonic media, where the interplay between spatial periodicity and temporal driving can be exploited to tailor mode selection, dispersion, and light–matter interaction within the cavity. Such capabilities further extend the accessible design space beyond that of static or purely electrically driven systems.

\section{Conclusions}
\label{Conclusions}

In summary, we demonstrate, through PL measurements, the coherent modulation of the THz QCL bandstructure, achieving an energy modulation amplitude of up to 10 meV at 8 GHz, induced by BAWs generated with an integrated transducer. The impact of BAW modulation on the bandstructure and optical gain has been investigated using simulations based on a phenomenological model within a quasi-adiabatic framework. In this approach, the strain-induced energy modulation is incorporated as a static contribution to the bandstructure Hamiltonian. For each discrete modulation amplitude, the simulations yield the steady-state response of the system, while the temporal evolution can be reconstructed by mapping these responses onto the corresponding phases of the acoustic cycle.
The coherent acoustic modulation of intersubband transitions in THz QCLs holds significant potential for expanding the functionality and application space of these devices. By introducing additional, mechanical degrees of freedom, this approach is set to enable dynamic control of gain, dispersion, and emission characteristics beyond what is accessible through conventional bias modulation, while decoupling the modulation mechanism itself from the electrical driving conditions. Such independent control opens pathways toward programmable THz sources, ultrafast waveform synthesis, and new regimes of frequency-comb operation. The next stages of development will require a systematic experimental assessment of BAW modulation on the optical properties and coherence of THz QCLs, together with the refinement of theoretical models capable of capturing the coupled electronic and strain dynamics. Ultimately, these advances are poised to establish mechanically driven QCLs as a versatile platform for coherent THz photonics.


\section*{Methods}

\textbf{THz QCL fabrication.}

The THz QCLs were fabricated using a single-plasmon waveguide geometry to keep fabrication complexity to a minimum and improve the reproducibility of the fabricated devices. The process began with photolithographic patterning to define the ridge waveguides, followed by wet chemical etching to form the laser mesas. A second lithography step was then used to define the top contact areas. Metal contacts, consisting of a Ni/AuGe/Au multilayer stack were deposited by evaporation. After metalization, the samples were annealed in a nitrogen atmosphere to improve ohmic contact formation.

\textbf{Bulk acoustic wave transducers.}

Bulk acoustic resonators (BARs)~\cite{Machado2019} were integrated on top of the QCL waveguide. A BAR consists of the bottom 200-nm-thick Au contact, 150-nm-thick piezoelectric ZnO film, a 20-nm-thick SiO$_2$ protective layer, and a top 10/30/10-nm-thick Ti/Al/Ti contact. In order to enable optical access to the QCL active region for PL measurements, both top and bottom contacts have lithographically defined openings. RF driving of the BAR around its resonance frequency generates a longitudinal bulk acoustic wave. The amplitude of the BAW can be precisely controlled by the radio-frequency power applied to the BAR. An RF probe was used to connect the bottom and top electrodes of the transducer to a RF generator with high-power output.

\textbf{Optical measurements with acoustic modulation.}

Optical measurements under acoustic excitation were carried out in a cold-finger liquid-He cryogenic probe station in the 5--10~K temperature range. A single-mode CW stabilized Ti-Sapphire laser tuned to 760 nm was used to non-resonantly excite the QCL. The laser was focused on the sample at normal incidence using a 10x objective resulting in a few-$\mu$m Gaussian-like spot positioned on the trap. PL measurements were carried out by projecting the PL image of the trap on the entrance slit of a single-grating spectrometer and recorded using a nitrogen-cooled CCD camera.

\section*{Data availability}
The measurement and numerical simulation data that support the findings within this study are included within the main text and Appendix and can also be made available upon a reasonable request from the corresponding author.

\bibliography{Mendeley_Bib}


\section*{Acknowledgments} 

We acknowledge the funding from DFG grant 359162958. 
The authors wish to thank W. Anders, N. Volkmer, C. Herrmann, D. Steffen, A. Riedel, and Dr. A. Tahraoui for sample preparation.
We thank Mr. M. Saeedi for the discussions about transducer fabrication.
The authors thank Dr. A. Hernández Mínguez for a critical review of the manuscript. 

\section*{Author contributions}

ASK performed the spectroscopic experiments with acoustic modulation and carried out the respective analysis. 
LS and XL developed the theoretical model.
VP performed the simulations and the related analysis of the results.
ASK and VP designed the lithography masks for acoustic transducers.
KB performed the MBE growth of the QCL. 
ASK has conceived the idea with the input from VP. 
ASK and VP have written the manuscript with the input from all authors.

\section*{Competing interests}
The authors declare no competing interests.



\section*{Appendix}

\subsection*{A1. PL spectrum of QCL active region}

\begin{figure}[!htb]
	\centering
		\includegraphics[width=0.45\textwidth, keepaspectratio=true]{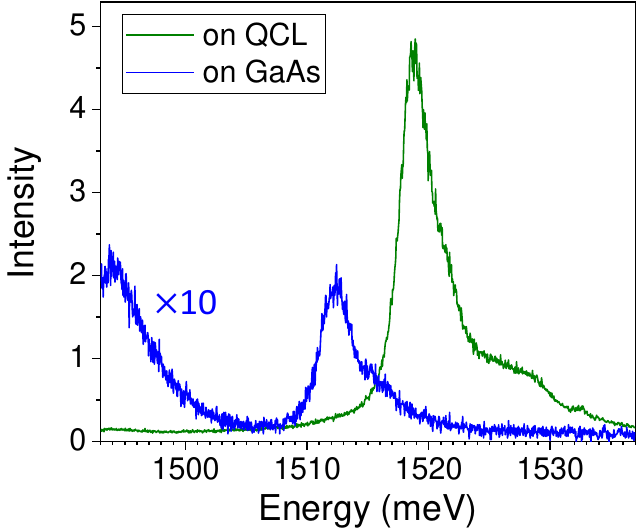}
		\caption{
            {\bf QCL PL spectrum.}
            Comparison of the PL of the GaAs substrate (blue curve) and the QCL region excited from the top (green curve). Note that the blue curve was multiplied by a factor of ten.
            }
	\label{FigSM-QCL-PL}
\end{figure}

Figure~\ref{FigSM-QCL-PL} compares the PL spectra measured from the bare GaAs substrate, i.e., a region of the wafer without the active region, and from the QCL active region, both excited at 760 nm and recorded at 10~K. As expected, the QCL PL peak is blueshifted with respect to that of the substrate.

\subsection*{A2. Pulsed optical and RF measurements}

\begin{figure}[!htb]
	\centering
		\includegraphics[width=0.45\textwidth, keepaspectratio=true]{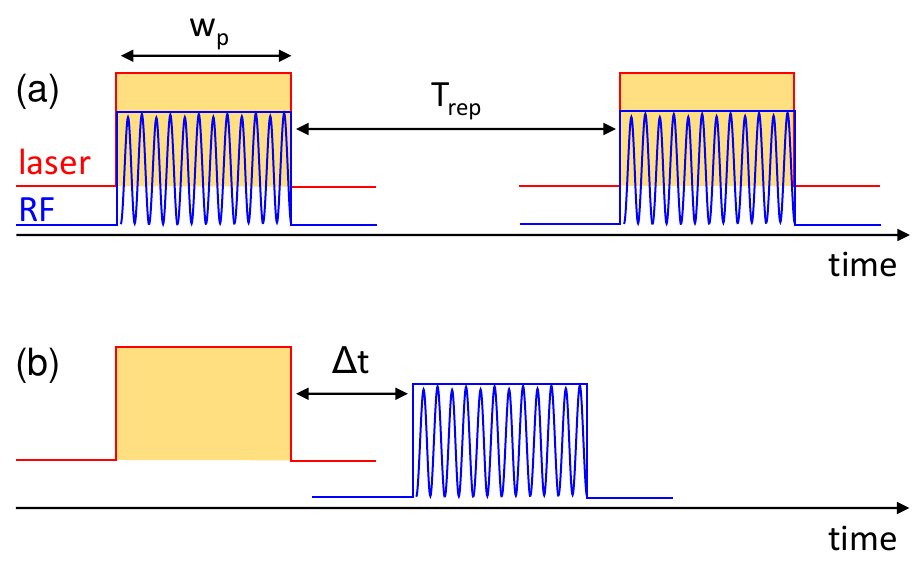}
		\caption{
            {\bf Pulsed experiment.}
            {\bf a} In-phase condition: temporal overlap between laser and RF pulses.
            {\bf b} Out-of-phase condition: the RF pulse is shifted by an amount of time exceeding the laser pulse width.
            }
	\label{FigSM-PulsedExperiment}
\end{figure}

Figure~\ref{FigSM-PulsedExperiment} shows the experimental scheme used for the modulation experiments. Here, the CW laser excitation was chopped (with a rotating blade) at a frequency $F_{\text{rep}} = 1/T_{\text{rep}} = 140$~Hz, where $T_{\text{rep}} \approx 7.1$~ms is the pulse period, producing pulses with temporal width of $w_{\text{p}} = 0.6$~ms. The pulse from the chopper was used to trigger the RF generator, which was set to generate RF pulse with the duration equal to $w_{\text{p}}$. The resulting low duty cycle $D = w_{\text{p}}/T_{\text{rep}} = 8.4 \%$ reduced the RF excitation-induced heating. Furthermore, to evaluate the effect of possible RF-heating experiments were run in two configurations: (i) in-phase, cf. Fig.~\ref{FigSM-PulsedExperiment}a and (ii) out-of-phase, by introducing a delay between the optical pulse and the RF one $\Delta t > w_{\text{p}}$, cf. Fig.~\ref{FigSM-PulsedExperiment}b. In the latter case, there was no temporal overlap between the laser and RF pulses.

\subsection*{A3. In-phase vs. out-of-phase modulation experiment}

\begin{figure}[!htb]
	\centering
		\includegraphics[width=0.45\textwidth, keepaspectratio=true]{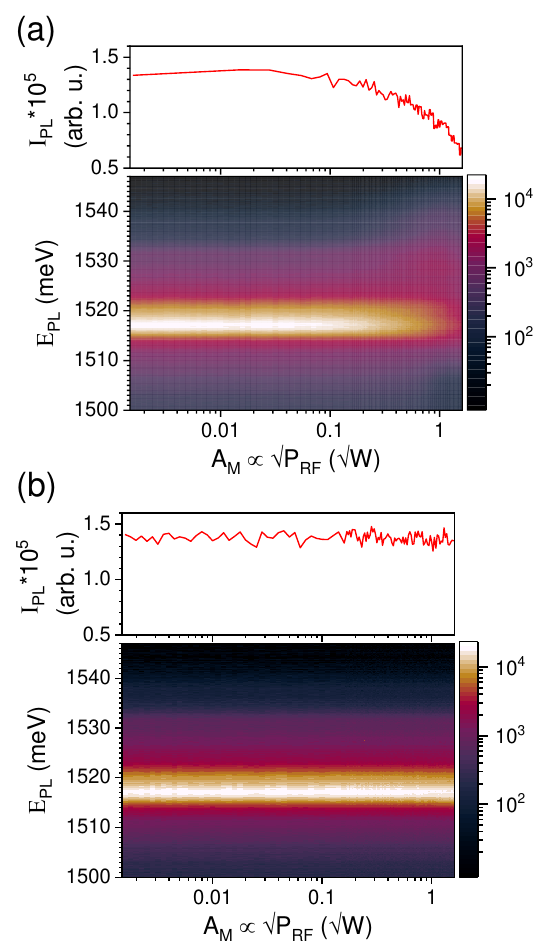}
		\caption{
            {\bf In-phase vs. out-of-phase modulation.}
            In-phase {\bf a} and out-of-phase {\bf b} maps showing the evolution of the QCL PL spectrum as a function of the RF amplitude ($\sqrt{P_{\text{RF}}}$) for RF frequency $\sim 7.9$~GHz. Upper insets in both panels show the PL intensity integrated around the main peak.
            }
	\label{FigSM-Phase-InAndOut}
\end{figure}

Figure~\ref{FigSM-Phase-InAndOut} shows a comparison between the in-phase (a) and out-of-phase (b) measurements of the QCL PL spectrum as a function of RF amplitude ($\sqrt{P_{\text{RF}}}$) for a fixed RF frequency of $\sim 7.9$~GHz. The procedure described in the Appendix A2 was used. In the out-of-phase case, the PL leaves both spectrum and intensity unchanged over the whole range of modulation amplitudes. This indicates a negligible effect of the RF-induced heating on the PL. In the case of the in-phase condition, both the spectrum and the integrated intensity change with increasing RF power. As discussed in the main text, the spectral modulation is due to the acoustic strain. We assume that the approx. $50 \%$ reduction of the PL intensity for the in-phase case may be related to the stray in-plane electrical fields that dissociate excitons, caused by the laterally propagating components of the BAW.  

\subsection*{A4. Gain response to sub-harmonic BAWs.}

Introducing BAWs with a periodicity larger than a single QCL period into active-region bandstructure simulations presents several challenges. On the one hand, explicitly simulating a full BAW period requires constructing a superperiod composed of multiple QCL periods in cascade. This approach often leads to numerical instabilities associated with the onset of quasi electric-field domain formation, i.e., non-uniform local electric fields across the structure. In the hybrid active-region designs considered here, this effect is further enhanced by the strong local charge accumulation associated with the injector-induced dipole. While resonant-phonon designs are less prone to such instabilities, their shorter period length would require BAW frequencies well beyond those achievable with ZnO-based BARs.

An alternative approach consists in partitioning the BAW modulation profile into segments, each mapped onto a single QCL period, and subsequently averaging the corresponding simulation results. However, this method introduces discontinuities in the potential profile at the boundaries of the simulated period, leading to unphysical transport and gain characteristics.

To overcome these limitations, we propose an approximate approach that enables estimation of the effect of sub-harmonic BAWs on QCLs without encountering the issues described above. For clarity, we present the method for the case of a BAW with twice the period of the active region, although the approach can be generalized provided that the approximation of the modulation profile remains valid for the intended application.

Starting from a modulation profile defined as 
$V(z) = M_{0} \cos\left(2 \pi z / (2w_\text{QCL}) + \phi\right)$, 
for which a whole rotation encompasses two QCL periods, each half-cycle can be approximated by a BAW with half the spatial period, a $\pi/2$ phase shift, and half the amplitude of the original modulation, reading
$V_{appr.}(z) = M_{0}/2*[1- \cos\left(2 \pi z/w_\text{QCL} + \phi+\pi/2\right)]$. 
Strictly speaking, the additional offset term only affects the absolute band-edge energy and does not influence the relative subband alignment relevant for transport and gain, so it can be safely neglected. Therefore, the approximated modulation profile reads
$V_{appr.}(z) = M_{0}/2*[\sin\left(2 \pi z/w_\text{QCL} + \phi\right)]$.
The total gain can then be estimated by averaging the results obtained from simulations with positive and negative modulation amplitudes $M_{0}$.

\begin{figure}[!htb]
	\centering
		\includegraphics[width=0.45\textwidth, keepaspectratio=true]{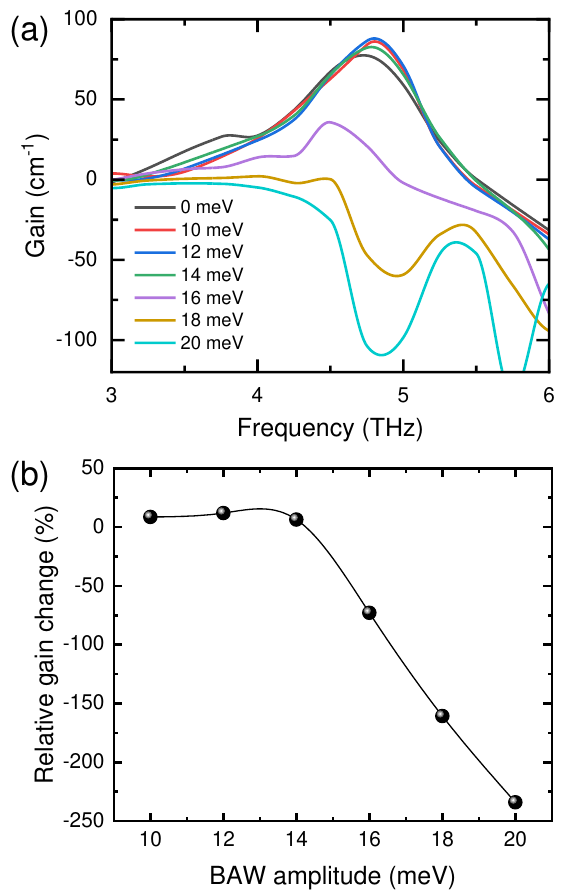}
		\caption{
            {\bf Gain modulation by sub-harmonic BAWs.}
           (a) Simulated optical gain spectra for an effective sub-harmonic BAW modulation with amplitudes ranging from 10 to 20 meV and twice the QCL spatial period, obtained using the approximate approach described in the text. The results correspond to the averaged response of two half-cycle modulations with opposite sign. The reference gain curve, corresponding to the unmodulated structure, is also provided as a reference. (b) Relative gain change at 4.75 THz with respect to the unmodulated case as a function of the BAW modulation amplitude. The gain remains nearly unchanged at lower amplitudes due to the partial cancellation between the two half-cycles, while a pronounced decrease is observed beyond a critical modulation strength where both phases significantly perturb the bandstructure.}

	\label{FigSM-Subharmonic}
\end{figure}

Figure~\ref{FigSM-Subharmonic}a shows the simulated gain curves for such an effective modulation with antinodes at the lasing wells of two consecutive periods, and amplitudes ranging from 10 to 20 meV. For amplitudes up to approximately 14 meV, the gain variations induced in the two half-cycles largely compensate each other, resulting in only minor net changes. The observed behavior is reminiscent of that obtained for a BAW at the fundamental spatial frequency, but with a reduced magnitude, consistent with the effective halving of the modulation strength. For larger amplitudes, a pronounced decrease in the gain is observed, again mirroring the behavior of the fundamental modulation, albeit at approximately twice the applied amplitude.

Figure~\ref{FigSM-Subharmonic}b reports the relative gain change at 4.75 THz with respect to the unmodulated case. The results clearly indicate that the gain remains nearly unchanged until the modulation amplitude becomes sufficiently large to significantly perturb the bandstructure in both positive and negative phases, thereby introducing losses in each case. Beyond this threshold, the gain decreases approximately linearly with increasing BAW amplitude.

\subsection*{A5. Characterization of a THz QCL from the investigated active region}

\begin{figure}[!htb]
	\centering
		\includegraphics[width=0.45\textwidth, keepaspectratio=true]{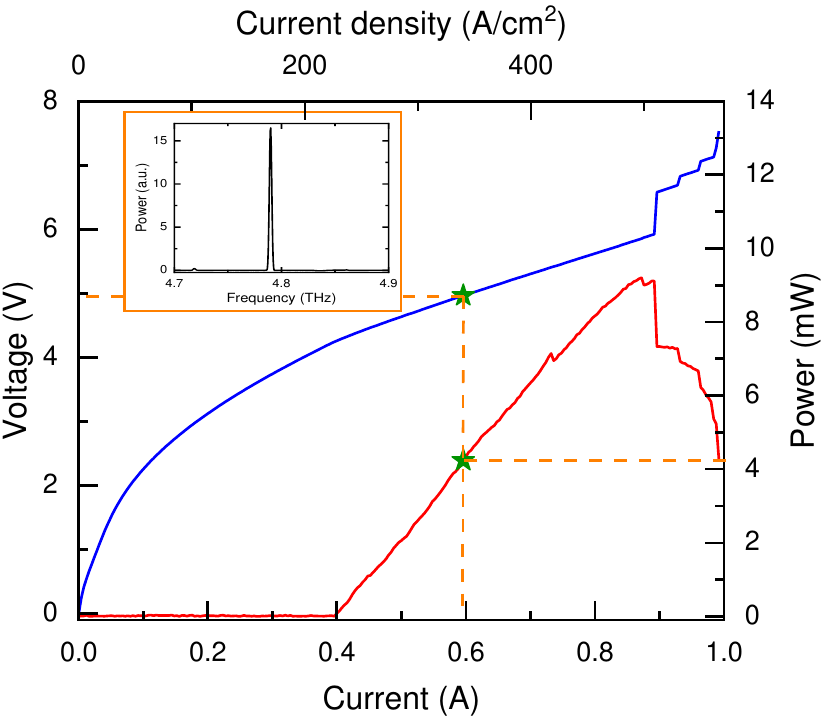}
		\caption{
            {\bf QCL LIV and spectral characterization.}
            $I$-$V$ (in blue) and $L$-$I$ (in red) curves recorded from a QCL realized from the active region employed in the PL experiments. The measurements have been performed at a heat sink temperature of 30 K and with CW bias. The star marks the bias condition corresponding to the field strength used in the simulations. Inset: emission spectrum of the QCL recorded with a Fourier transform infrared spectrometer at 50 K. 
            }
	\label{FigSM-QCL-IV}
\end{figure}

Figure~\ref{FigSM-QCL-IV} presents the $L$-$I$-$V$ characteristics of a 1460 µm × 120 µm × 10.8 µm (length, width, and thickness of the active region, respectively) QCL fabricated by cleaving a waveguide containing the active region investigated in this work. No BAR was integrated on top of this part of the waveguide, and electrical connections were realized using standard Au wire bonding. The device was tested in CW mode at a heatsink temperature of 30 K and exhibits a lasing threshold at 392 mA (224 A/cm²), corresponding to -4.23 V.

The simulations were performed at an electric field of -4.6 kV/cm, corresponding to approximately 4.97 V and 596 mA (340 A/cm²). At this operating point, the device delivers 4.23 mW of output power. This condition, indicated by a star symbol in Fig.~\ref{FigSM-QCL-IV}, represents a conservative intermediate choice between threshold and the maximum operating bias achievable with this active region. The inset shows the spectrum of the QCL, which presents a single mode at 4.79 THz, measured with a Fourier transform infrared spectrometer at 50 K.

\subsection*{A6. BAW propagation}

\begin{figure}[!htb]
	\centering
		\includegraphics[width=0.45\textwidth, keepaspectratio=true]{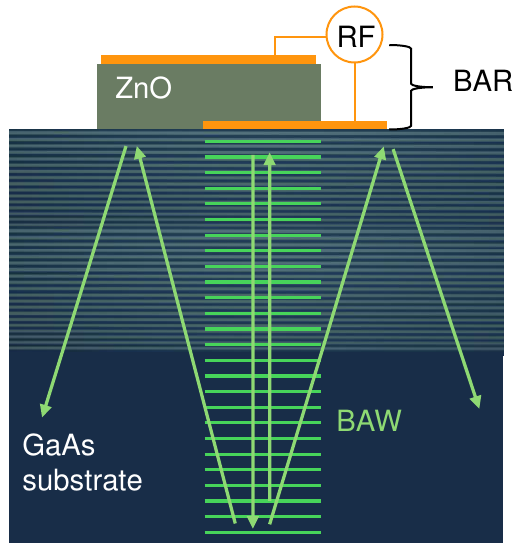}
		\caption{
            {\bf BAW propagation.}
            A schematic showing the approximate propagation of a BAW excited by an RF-driven BAR. The active region of the BAR is given by the overlap of the top and the bottom contacts (orange shapes). The active region emits a BAW, whose wavefronts are depicted by the horizontal green lines. Due to the edge effects, the BAW spreads laterally, which leads to the lateral propagation, which is enhanced by multiple reflections between the top and bottom smooth surfaces of the sample, as depicted by the diagonal green arrows.
            }
	\label{FigSM-BAW-Propagation}
\end{figure}

The sketch of Fig.~\ref{FigSM-BAW-Propagation} depicts the propagation path of the BAW excited by a BAR. Here the active region of the transducer is defined by the overlap of the top and bottom electrodes. The BAW propagates mostly perpendicular to the BAR surface. 

However, close to the edges of the active area, a slightly diverging BAW is generated, which can propagate large lateral distances due to the multiple reflection between the top and bottom smooth surfaces of the sample, especially at low temperatures, where the acoustic absorption is small. This phenomenon has been established and used recently to modulate microcavity exciton-polaritons at GHz frequencies~\cite{Kuznetsov2021,Kuznetsov2023}. 

\end{document}